\documentstyle[twocolumn,prl,aps]{revtex}
\begin{document}
\draft
\input{psfig}
\preprint{TO BE SUBMITTED TO SCIENCE!}
% FOR TWO COLUMN  ACTIVATE THE LINE BELOW 
\twocolumn[\hsize\textwidth\columnwidth\hsize\csname @twocolumnfalse\endcsname
\title{Giant anharmonicity and non-linear electron-phonon coupling in MgB$_{2}$; \\
Combined first-principles calculations and neutron scattering study}
\author{T. Yildirim$^{(1)}$, O. G\"{u}lseren$^{(1,2)}$, J. W. Lynn$^{(1)}$, 
C. M. Brown$^{(1,3)}$, T. J. Udovic$^{(1)}$, 
H. Z. Qing$^{(1,3)}$,N. Rogado$^{(4)}$, 
K.A. Regan$^{(4)}$, M.A. Hayward$^{(4)}$, J.S. Slusky$^{(4)}$, 
T. He$^{(4)}$, M.K. Haas$^{(4)}$, P.  Khalifah$^{(4)}$, 
K. Inumaru$^{(4)}$, and R.J. Cava$^{(4)}$}
\address{$^{(1)}$ NIST  Center for Neutron Research, National Institute of Standards
and Technology, Gaithersburg, MD 20899 }
\address{$^{(2)}$ Department of Materials Science  and Engineering, 
University of Pennsylvania, Philadelphia, PA 19104}
\address{$^{(3)}$ University of Maryland, College Park, MD }
\address{$^{(4)}$  Department of Chemistry and Princeton Materials Institute,
Princeton University, Princeton, NJ 08544 }
\date{\today}
\preprint{Submitted to Phys. Rev. Lett.}
\maketitle

\begin{abstract}
We report first-principles calculations of the electronic band structure and lattice
dynamics for the new superconductor MgB$_{2}$. The excellent agreement between
theory and our inelastic neutron scattering measurements of the phonon density of
states gives confidence that the calculations provide a sound description of the
physical properties of the system. The numerical results reveal that the in-plane
boron phonons (with E$_{2g}$ symmetry) near the zone-center are very anharmonic,
and are strongly coupled to the partially occupied planar B $\sigma$ bands near the
Fermi level. This giant anharmonicity and  non-linear electron-phonon coupling is 
key to quantitatively explaining the observed high T$_{c}$ and  boron isotope 
effect in MgB$_{2}$
\end{abstract}

%%%
%63.20.Ry   Anharmonic lattice modes  
%63.20.Kr   Phonon–electron and phonon–phonon interactions  
%61.12.-q   Neutron diffraction and scattering 
%74.25.Jb   Electronic structure  
%74.25.Kc   Phonons 


]

\bigskip
The recent discovery of superconductivity at 40~K in the MgB$_{2}$ binary alloy
system\cite{1} has triggered enormous interest in the structural and
electronic properties of this class of materials. The system has a very
simple crystal structure\cite{2}, where the boron atoms form 
graphite-like sheets separated by hexagonal layers of Mg atoms (see inset
to Fig.~1). Our pseudopotential plane wave band structure calculations show
that the bands near the Fermi level arise mainly from the $p_{x,y}$ $\sigma$
bonding orbitals of boron, while the Mg does not contribute appreciably to 
the conductivity, in good agreement with initial reports from other 
groups\cite{3,4,anderson}. In the case of graphite these $\sigma$ bands
are full, but for MgB$_{2}$ they are partially unoccupied, creating
a hole-type\cite{5} conduction band like the high-T$_{c}$ cuprates.
In contrast to the cuprates, however, the normal-state conductivity is
three-dimensional in nature instead of being highly anisotropic, thus
eliminating the ``weak-link" problem that has plagued widespread 
commercialization of the cuprates. The normal-state conductivity\cite{5,6,7}
is also one to two orders-of-magnitude higher than either the Nb-based
alloys or Bi-based cuprates used in present day wires, and this feature
combined with low cost and easy fabrication\cite{7,8} could
make this class of materials quite attractive for applications.

From a fundamental point of view the central question is whether
the high T$_{c}$ in this new system can be understood within the 
framework of a conventional electron-phonon mechanism, or a more
exotic mechanism is responsible for the superconducting pairing. 
The observed boron isotope effect\cite{9} argues for an electron-phonon
mechanism, while the positive Hall coefficient\cite{5} suggests
similarities with the cuprates\cite{10}. To answer this question,
we have carried out inelastic neutron scattering measurements of the
phonon density of states, and compare these results with detailed
first-principles calculations of the lattice dynamics (and electronic)
calculations for MgB$_{2}$. Excellent agreement is found between
theory and experiment. More importantly, the numerical results
demonstrate that the in-plane boron phonons near the zone-center
(with E$_{2g}$ symmetry at $\Gamma$) are very anharmonic and
strongly coupled to the partially occupied conduction band (planar 
B $\sigma$ bands)
near the Fermi level, providing the large electron-phonon
interaction in this system. 
Due to these strong  anharmonic phonon modes, the electron-phonon
coupling is non-linear, providing the  essential ingredient to 
explain the high T$_{c}$ and  boron isotope effect in MgB$_{2}$.

The
results of the neutron measurements and calculations for the phonon density of
states are summarized in Fig.~1.  For the experimental data, the 5g
polycrystalline sample was prepared in the usual way with the $^{11}$B isotope to
avoid neutron absorption problems\cite{11}, and was characterized by magnetization,
x-ray, and neutron diffraction.  Inelastic neutron scattering measurements were
made to determine the generalized phonon density of states (GPDOS), which is
the phonon density of states weighted by the cross section divided by the mass
of each atom.  The data were collected from 7 K to 325~K on the Filter Analyzer
Spectrometer in the range 5-130 meV, and on the Fermi chopper instrument for
energies of 0.5-30 meV.  Additional details of the experiment and analysis can
be found elsewhere\cite{12,13}.  The experimental data indicate two bands of
phonons, one below 40 meV corresponding primarily to the acoustic phonon modes,
and one above 50 meV that mostly involves the boron motions.  The high-energy
portion of the spectrum shows a sharp cutoff at about 100 meV.  
The data are in basic agreement with the results of Osborn {\it et al.}\cite{osborn},
although the present data were obtained with better overall energy resolution. Our
results do not agree with Sato {\it et al.}\cite{sato}, who reported a feature
at 17 meV in  the GPDOS. We did not observe this feature at any temperature and conclude
that it is not intrinsic to the GPDOS. We also have measured the temperature 
dependence of the spectrum and find
no substantial changes in the basic features  from 7 K up to 200~K, 
while a very modest softening of some of the modes was observed in going to
325~K.

\bigskip
\centerline{\psfig{figure=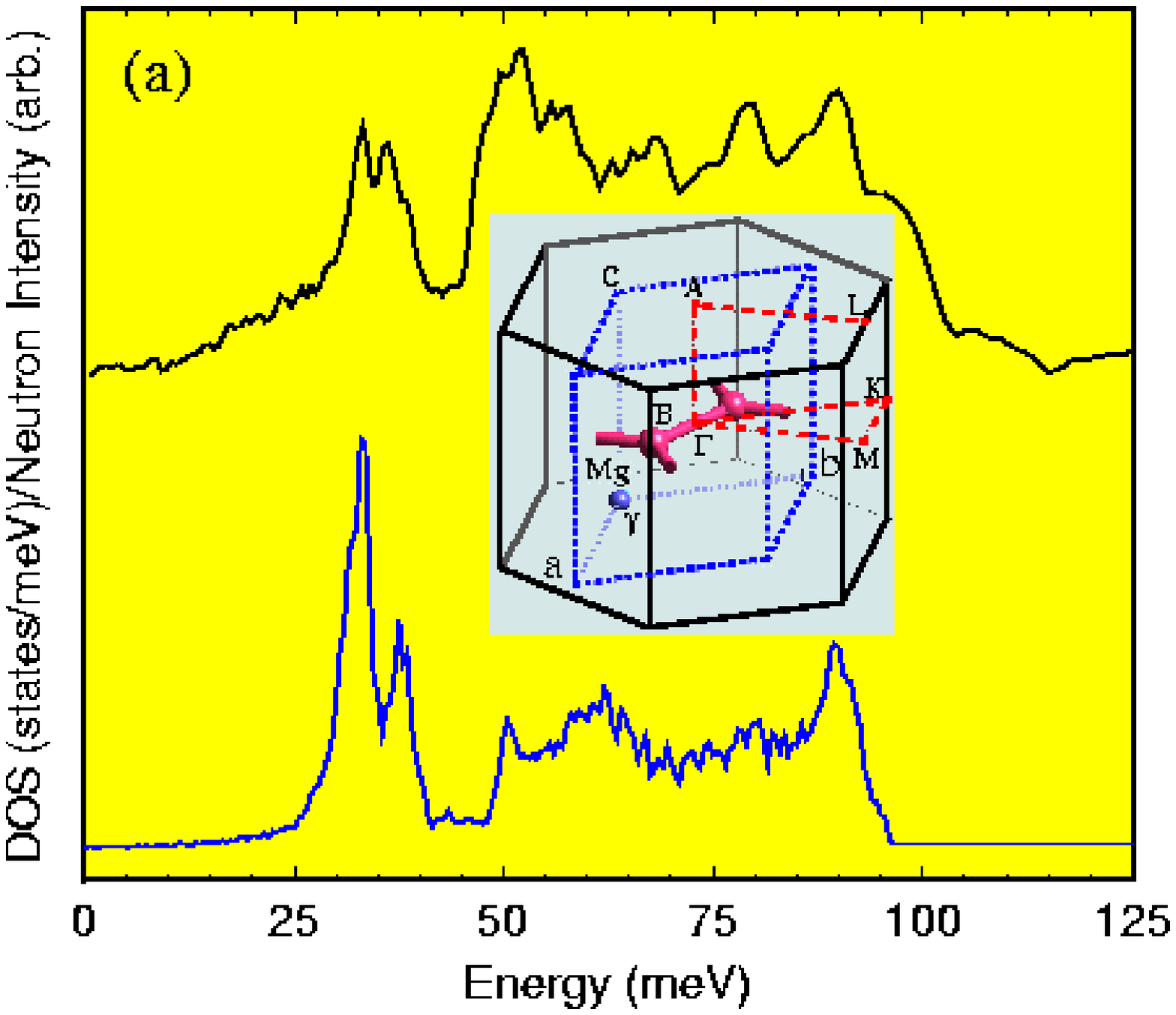,width=75mm}}

\centerline{\psfig{figure=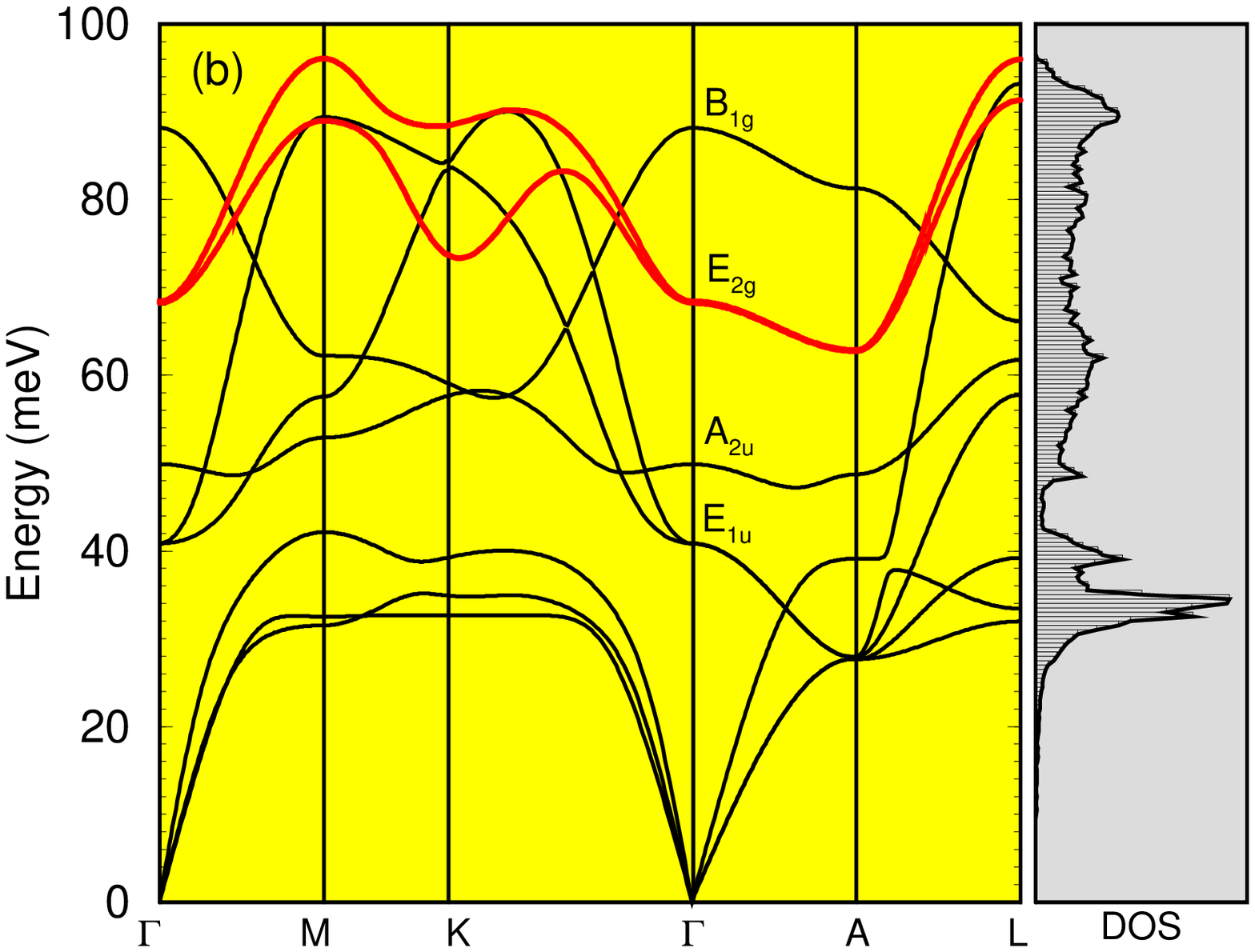,width=75mm}}

{\bf Fig.~1} {\small 
(a) Generalized phonon density of states for MgB$_2$ (top
curve) determined from inelastic neutron scattering measurements, and the
calculated phonon density of states (DOS) (bottom curve). The inset shows the
unit cell of MgB$_2$ along with the Brillouin zone and the high-symmetry
directions. (b) Calculated dispersion curves of all phonons and the
corresponding DOS (right panel).  At the zone center ($\Gamma$),
 the symmetries of the
modes are also indicated. The modes indicated by the red-line dominate the
electron-phonon coupling.
}

\bigskip

In order to understand the origin of the features
observed in the neutron data, we have carried out detailed first-principles
calculations of the electronic band structure and lattice dynamics.  The
calculations were performed using the pseudopotential plane wave method\cite{14},
within the generalized gradient approximation\cite{gga}.  
We employed plane waves with an
energy cutoff of 500 eV, and the ultra-soft pseudopotentials for Mg and B\cite{usp}.  
The total energy and forces converged within 0.5 meV/atom and 0.01 eV/\AA,
respectively. Brillouin-zone integrations\cite{kpts} were carried out using 
$dk = 0.02$ \AA$^{-1}$,
generating $15\times 15\times 11$ and $9\times 9\times 6$ k-points for 
$1\times 1\times 1$ and $2\times 2\times 2$ supercells,
respectively.  The lattice parameters a and c were relaxed before the lattice
dynamics calculations, and the force constant matrix was obtained by the
direct-force method using periodically repeated supercells\cite{16}. The optimized
lattice parameters are in good agreement with our  experimental values within
1\%. The calculated density of states obtained from these calculations is shown
in Fig.~1a, and we see that the agreement between experiment and theory for the
energies of the modes is excellent.  The agreement for the intensities is also
good considering that the observed spectrum is a weighted density of states
rather than the actual DOS, and the wave vector averaging may include some
coherency effects\cite{12}.  
The calculated phonon dispersion curves along the
high-symmetry directions of the Brillouin-zone (see inset to Fig.~1) are shown
in Fig.~1b.  All the modes are found to be quite dispersive as expected for the
small size of the unit cell, but divide nearly completely into acoustic (below
$\approx 40$ meV) and optical (50-100 meV) bands, with the only exception being along
the A-L direction.  All the optical modes are weakly dispersive along the $\Gamma$-A
direction, reflecting the layered nature of the MgB$_2$ crystal structure.

There are four distinct phonon modes at the zone center $\Gamma$, 
as indicated in Fig.~1b.  The A$_{2u}$ and B$_{1g}$ singly-degenerate 
modes involve only vibrations along the
c axis; for B$_{1g}$ the boron atoms move in opposite directions while the 
Mg ion is
stationary, while for the A$_{2u}$ mode the Mg and B planes move in opposite
directions along the z-axis.  The other two modes are doubly degenerate and
involve only in-plane motions (along the x or y axes).  For the E$_{1u}$ mode the Mg
and B planes vibrate in opposite directions along the x (or y) axis.  The
calculated energies of all zone-center phonons are listed in Fig.~2, and for the three
types of modes just discussed there is good agreement between  
calculations
reported by other groups\cite{3,4,anderson}.
%%%; Raman and infrared measurements have not been reported yet.  
For the E$_{2g}$ mode, however, there are large discrepancies between
the results reported for various calculations.  The boron ions for this mode
vibrate in opposite directions along the x (or y) axis, with the Mg ions
stationary.    The discrepancies noted above are resolved when the strong
anharmonicity associated with the in-plane displacements of the boron atoms is
taken into account. 

To further
investigate this behavior, we plot in Fig.~2 the energy  as the atoms are
displaced by an amount $u$ according to each of the four eigenmodes.  For the
B$_{1g}$, A$_{1u}$, and E$_{1u}$ modes the distortion energy can be very 
well approximated by
the harmonic expression $E(u)= A_2 u^2$  up to $u/a = 0.065$.  It is unusual that the
phonons remain harmonic at these large distortions, but this is consistent with
the fact that we did not observe any substantial changes in the measured GPDOS
with temperature.

 \bigskip
\centerline{\psfig{figure=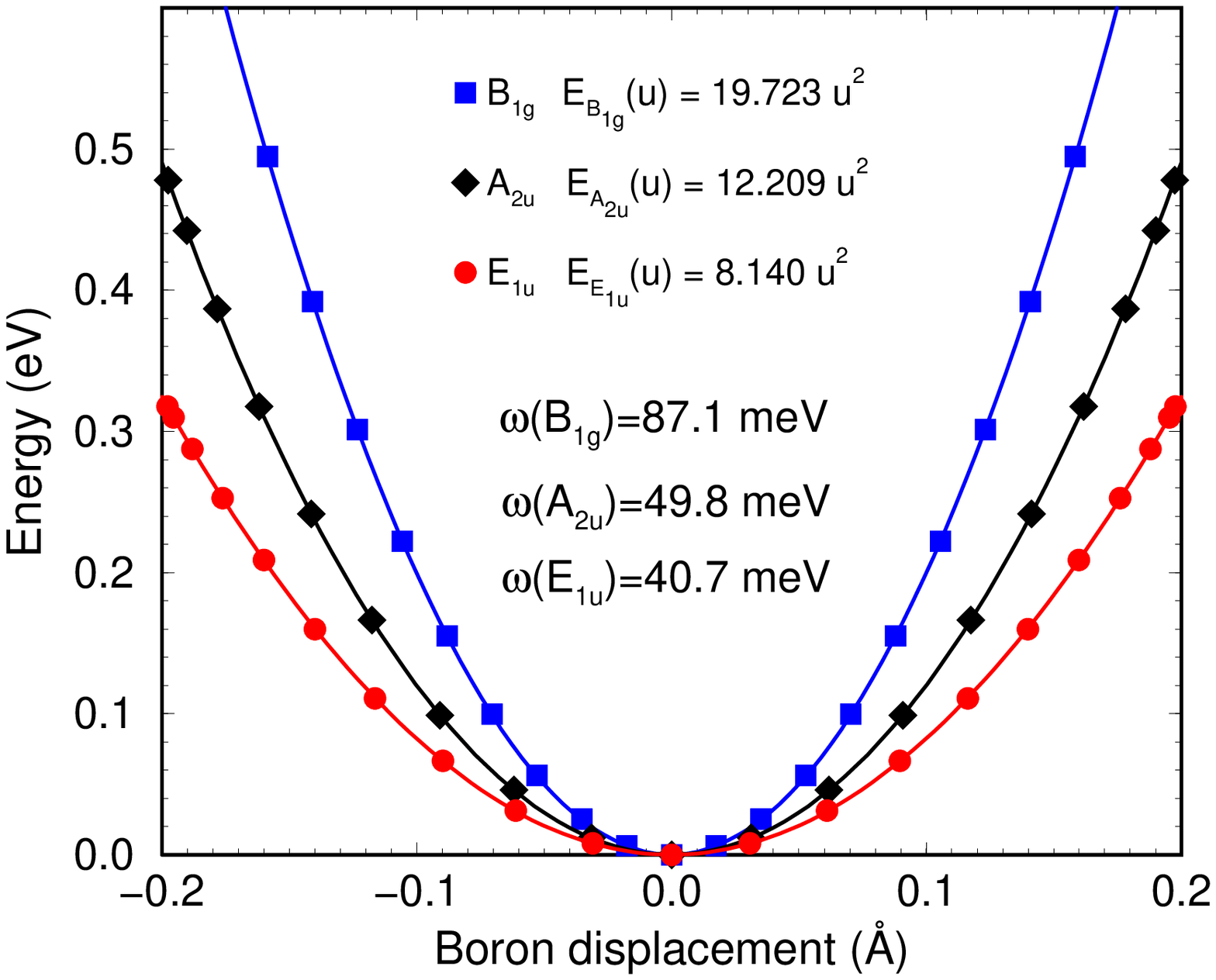,width=65mm}}

\centerline{\psfig{figure=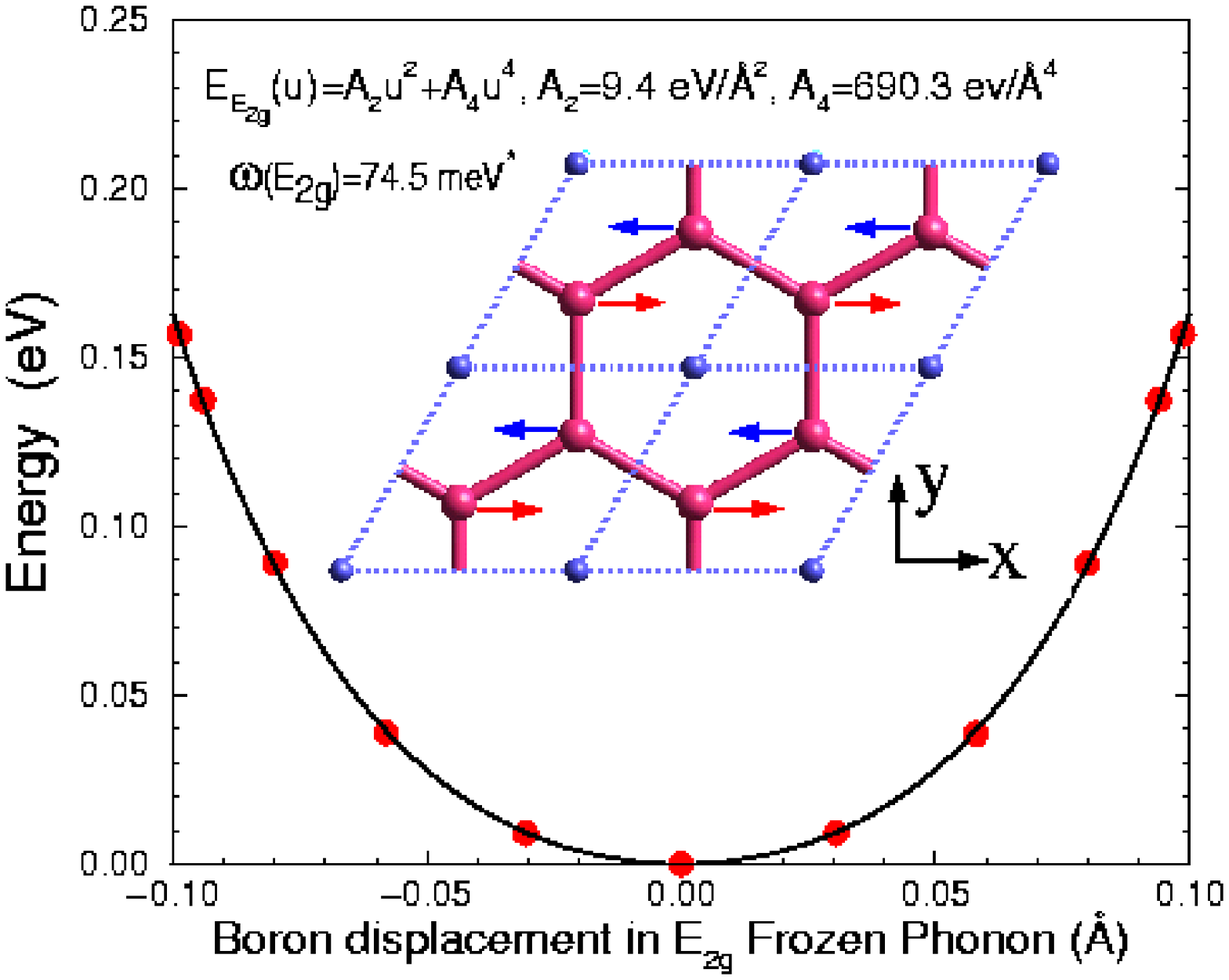,width=65mm}}

{\bf Fig.~2} {\small 
Energy
curves as a function of boron displacement for each zone center frozen-phonon
mode. The top panel shows that the B$_{1g}$, A$_{2u}$, and 
E$_{1u}$ modes are harmonic up to
$ u/a \approx 0.065$.  The bottom panel shows that the E$_{2g}$ boron 
in-plane mode is very
anharmonic.  The inset indicates one of the E$_{2g}$ modes.  Animations
of these modes and their coupling with the electronic structure can
be found at {\bf http://www.ncnr.nist.gov/staff/taner/mgb2}.
}
\bigskip

\noindent
The most interesting observation, though, is the totally
opposite behavior of the E$_{2g}$ in-plane boron mode, as shown in Fig.~2b.  
The potential well is very shallow at small displacements 
and increases rapidly at
large displacements, and can be fit very well to $E(u)=A_2 u^2 + A_4 u^4 $
with a  large ratio of$ A_4/A_2^2 \approx 8$.  
This indicates that this mode is unusually
anharmonic and that we are in a non-perturbative regime.  Hence the estimated
harmonic energy, $\omega_H$(E$_{2g}$) = 60.3 meV will be lower than the actual energy. 
Using
Hartree-Fock decoupling, one gets $E(u)= (A^2 + 3 A_4 <u^2> ) u^2$, 
where $<u^2> = /2M\omega_{sch}$ and $\omega_{sch}$ is the 
Self Consistent Harmonic (SCH)\cite{sch} solution of $\omega^2_{sch} = [(A_2
+ 3 A_4 <u^2>)/M]^{1/2}$.  This yields a better estimate of 
$\omega_{sch}$ $\approx 70$ meV, 
a 17\% enhancement of the
harmonic phonon energy $\omega_H(E_{2g})$.  We can calculate the {\it exact}
 energy levels of the
potential by numerically solving the 
Schrodinger equation and obtain 
$\omega(E_{2g}) =74.5 $ meV, a 25\% enhancement of the harmonic value 
and also the T$_C$, which then
matches with the peak in the GPDOS. 
%Since the E$_{2g}$ mode is Raman active,
%measurement of this mode by Raman scattering will be an important test of our
%prediction.

Since the in-plane motions of the boron will change the
boron orbital overlap, significant electron-phonon coupling can be expected for
the planer $\sigma$  boron conduction band at the Fermi level and this  plays an
important role in the superconducting pairing.
This can be most easily seen by comparing
the  band structure of the undistorted lattice to that of  
distorted one by a  zone center phonon. 
For the harmonic  phonons we did not see any
significant changes near the Fermi level. However, for
E$_{2g}$ modes, there is a significant splitting of the
bands as well as shifts near the Fermi level as shown in 
Fig.~3, indicating strong electron-phonon coupling. 
%%%(which is in complete agreement with the other reports\cite{3,4}). 

\bigskip
\centerline{\psfig{figure=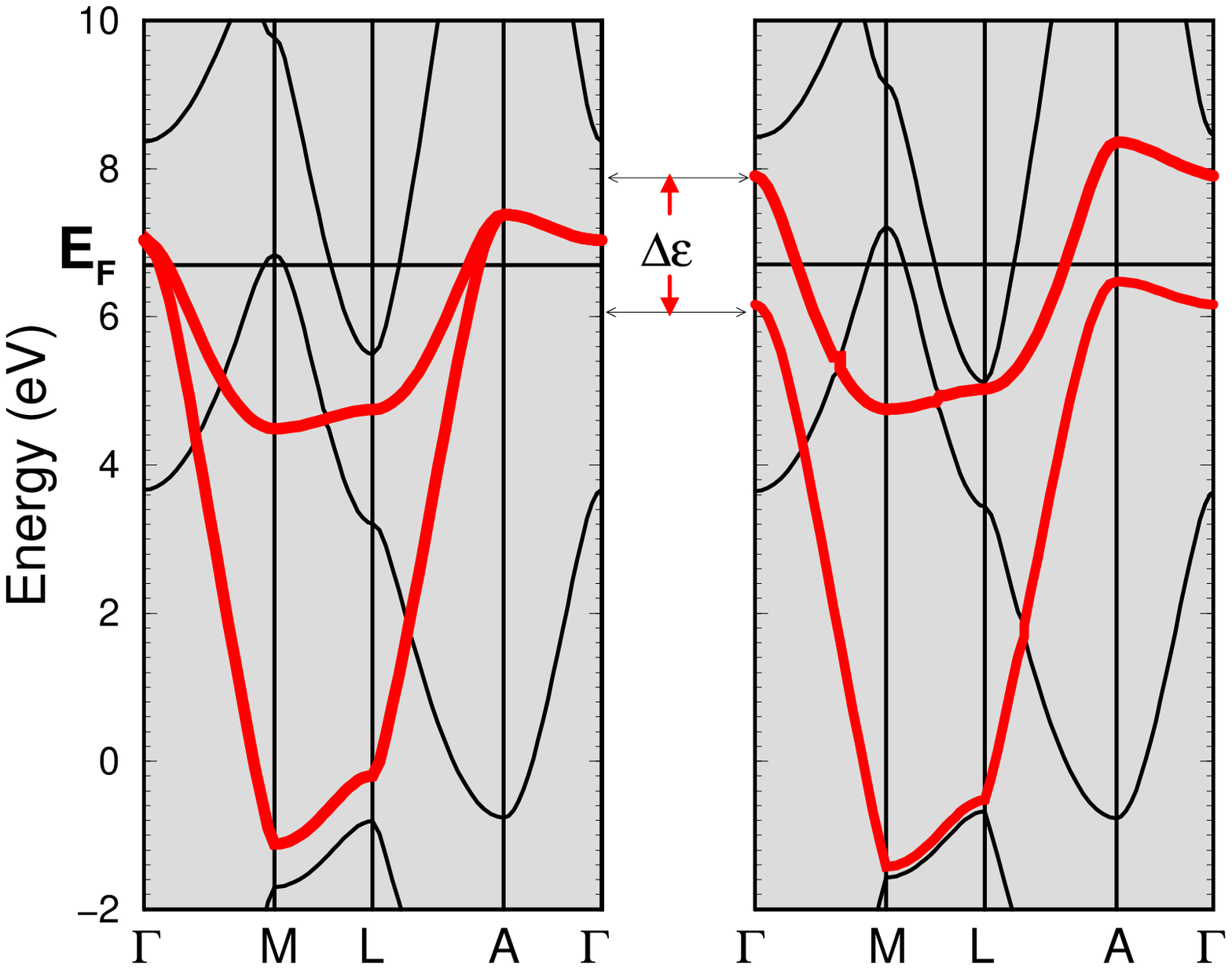,width=55mm}}

{\bf Fig.~3} {\small Band structure of the undistorted (left)
and distorted structures (right) by E$_{2g}$ phonons ($u_{B} \approx 0.06 $\AA). 
Band structure of the other three frozen-phonon structures do not show
any significant changes. }

\bigskip

To determine the electron-phonon (EP) coupling quantitatively,
we evaluated the Fermi-surface averaged deformation potential,
\begin{equation}
\Delta = \big< [\delta \epsilon ({\bf k}) - \delta\mu ]^{2} \big>,
\end{equation}
for each zone-center frozen phonon\cite{16}.  
Here $\delta \epsilon ({\bf k})$ is the change in the one-particle energy with
momentum ${\bf k}$ due to the frozen phonon, $\delta \mu$ the corresponding
change in the chemical potential. $< >$ denotes an average of ${\bf k}$ over the Fermi
surface, which we have carried out numerically.

For the harmonic B$_{1g}$, A$_{2u}$, and E$_{1u}$ 
phonons we calculated  an insignificant
coupling, and conclude that the electron-phonon coupling is
negligible for these modes. For the E$_{2g}$ modes, Fig.~4a shows that 
the coupling $\Delta (u)$ is large and has
both quadratic and quartic terms in frozen-phonon amplitude, $u$,
indicating significant non-linear electron-phonon coupling.  
Neglecting non-linear cross
terms, the electron-phonon coupling constant  $\lambda$
for our  anharmonic phonon is  approximately given by[17]

\begin{equation} 
\begin{small}
\lambda = N(E_{F}) \left( B'_{2}  \sum_{n} \frac{|<n|Q|0>|^{2}}{E_{n} -E_{0}} 
+ B'_{4}  \sum_{n} \frac{|<n|Q^{2}|0>|^{2}}{E_{n} -E_{0}} \right),
\end{small}
\end{equation}
where $N(E_{F})$ is the
total density of states at the Fermi energy and equal to 0.69 states/eV/cell. 
$E_n$
and $|n>$ are the eigenvalues and eigenfunctions of the oscillator E$_{2g}$ 
in its adiabatic potential shown in Fig.~2b. 
$B'_2$ and $B'_4$ are the first and second-order
EP coupling, respectively, and obtained from expansion of 
$\Delta(u)$. $Q$ is the normal coordinate which is related to 
boron displacement $u$ via a normalization constant. We calculate
a  total electron-phonon coupling $\lambda = 0.907 (0.922)$ for 
M$_{B}$=10 (11). 
Using the McMillan expression for T$_C$\cite{18}
and taking a typical value for $\mu^{*} = 0.15$, 
we obtain a T$_C$  of 39.4~K and 38.6~K for M$_{B}$=10
and M$_{B}$=11, respectively, yielding an isotope effect
$\alpha = 0.21$\cite{taniso} (see Fig.~4b). 
These estimates are  in excellent agreement
with experiments. 
  Within the above approximation\cite{17}, the non-linear
EP coupling increases T$_C$ by about 10\%. Our preliminary calculations of 
$\lambda$
including other non zone-center phonons indicates that the numbers will not
change significantly. Therefore we conclude that MgB$_2$ is a conventional
electron-phonon superconductor in the strong-coupling regime. 

\bigskip
\centerline{\psfig{figure=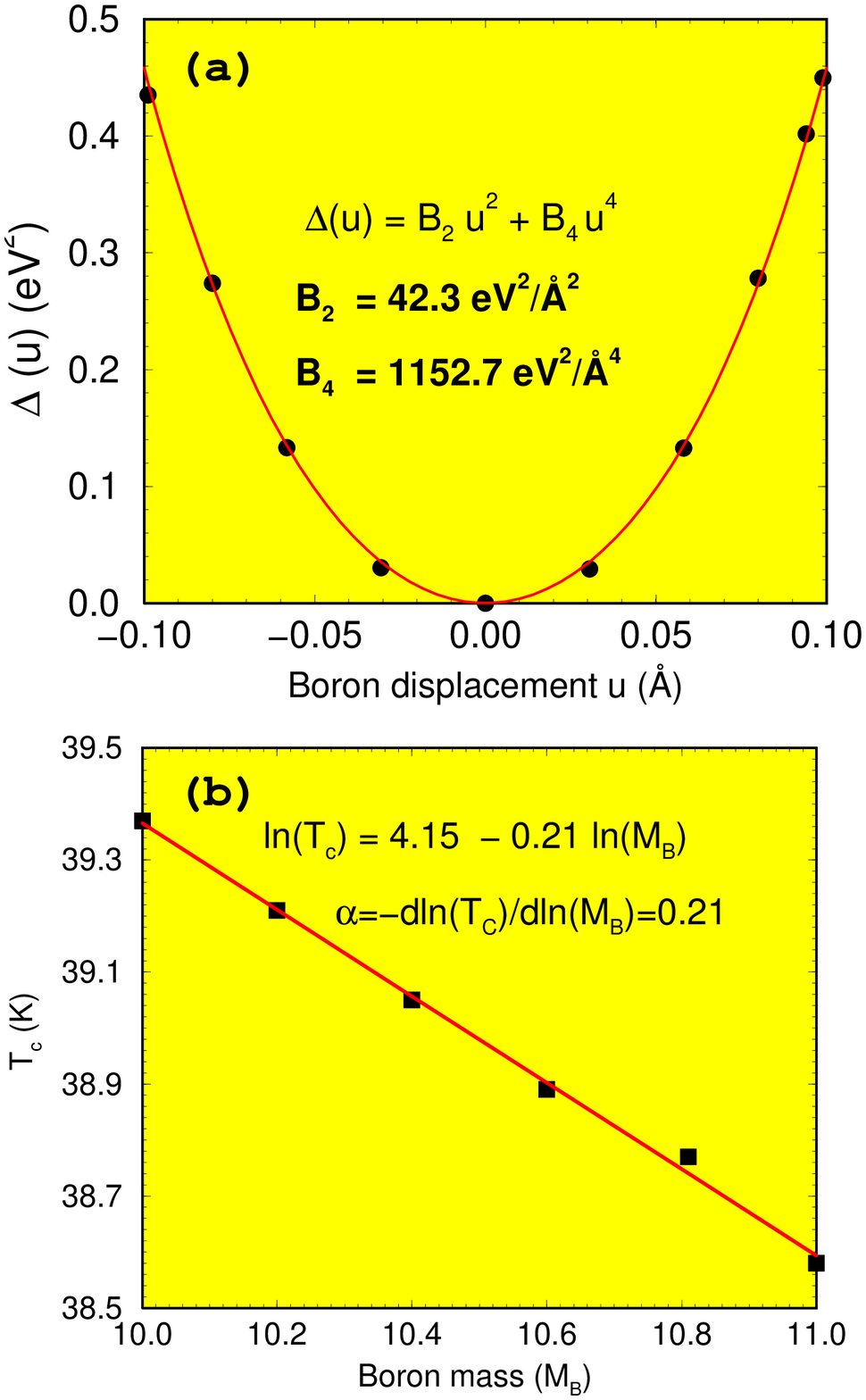,width=60mm}}

{\bf Fig.~4} {\small 
(a) Calculated 
Fermi-surface averaged deformation potential $\Delta (u)$. 
The large quartic term indicates significant non-linear electron-phonon
coupling.
(b) T$_{c}$ versus boron mass, indicating a boron isotope effect
$\alpha=0.21$, significantly reduced from harmonic BCS value of 0.5
due to giant anharmonicity\cite{taniso}. Within our theory, we expect zero
isotope effect for Mg, in good agreement with recent report of
$\alpha_{Mg}$=0.02\cite{mgiso}.
}

 It is
interesting to note that there seems to be a close correlation between the
anharmonicity of the E$_{2g}$ in-plane modes and the B-B bond length (d$_{BB}$). 
For MgB$_{2}$,
d$_{BB}$ is 1.764 \AA, significantly stretched from its optimal value of 
1.65 \AA $\;\;$ in elemental planar boron, probably due to repulsive 
interactions between the Mg
and B ions. This explains the unusual anharmonicity and observed high T$_C$ in
MgB$_{2}$. 
%On the other hand, for BeB$_2$ the optimized d$_{BB}$ is found to 
% be 1.67 \AA $\;\;$ with no significant anharmonicity. It is interesting 
%that this compound does not superconduct even though its band structure 
%is quite similar to that of MgB$_2$.
For hypothetical CaB$_2$, the calculated d$_{BB}$ is quite large (1.84 \AA) and we
calculate very large anharmonicity, which may explain the instability of CaB$_2$.
It then appears  likely that MgB$_2$ is fortuitously just at the phase boundary.
It will be interesting to see if it is possible to expand the d$_{BB}$ in 
MgB$_2$ by Ca substitution on the Mg site to increase T$_{C}$ further.

\bigskip
\noindent
{\bf NOTE ADDED IN PROOF:}
After submission of our work, a recent Raman study\cite{raman}
has reported 77 meV for the E$_{2g}$ mode with a very large width, 
confirming our predictions.

\noindent
{\bf ACKNOWLEDGEMENTS:}  
We acknowledge many useful discussion with 
R. L.  Cappelletti and  D. A. Neumann.

\end{document}